\documentclass{article}
\usepackage{latexsym}
\usepackage{url}
\usepackage{sss}
\usepackage[pdftex]{graphicx} 
\usepackage{amsmath,amssymb}
\usepackage[numbers]{natbib}
\usepackage{flushend}
\begin{document}
\date{}
\title{\LARGE{\bf 	Statistical Properties of Car Following: \\ Theory and Driving Simulator Experiments}
      }
\author{
	Hiromasa Ando, Ihor Lubashevsky*, Arkady Zgonnikov, Yoshiaki Saito 
	\\
	University of Aizu, Tsuruga, Ikki-machi, Aizu-Wakamatsu City, Fukushima 965-8560, Japan
	\\ 
	*E-mail for correspondence: i-lubash@u-aizu.ac.jp 
}

\maketitle
\thispagestyle{empty}

\abstract{

A fair simple car driving simulator was created based on the open source engine TORCS and used in car-following experiments aimed at studying the basic features of human behavior in car driving.  Four subjects with different skill in driving real cars participated in these experiments. The subjects were instructed to drive a car without overtaking and losing sight of a lead car driven by computer at a fixed speed. Based on the collected data the distributions of the headway distance, the car velocity, acceleration, and jerk are constructed and compared with the available experimental data for the real traffic flow. A new model for the car-following is proposed to capture the found properties. As the main result, we draw a conclusion that human actions in car driving should be categorized as  generalized intermittent control with noise-driven activation. Besides, we hypothesize that the car jerk together with the car acceleration are additional phase variables required for describing the dynamics of car motion governed by human drivers. 
}

\section{Introduction}

According to the modern point of view  \cite{gawthrop2011intermittent,loram2011human,balasubramaniam2013control,milton2013intermittent,asai2013learning} human control over unstable mechanical systems should be categorized as intermittent. In this context the intermittency implies discontinuous control which repeatedly switches off and on instead of being always active throughout the process. As a result, the actions of a human operator in controlling a mechanical system form a sequence of alternate fragments of his passive and active behavior. According to the current state of the art, this type control being rather efficient on its own is a natural consequence of human physiology \cite{loram2011human}. 

The concept of event-driven intermittency is one of the most promising approaches to describing human control. The threshold-driven activation is widely accepted as the basic mechanism of such intermittent control. It posits that the control is activated when the discrepancy between the goal and the actual system state exceeds a certain threshold. Models based on the notion of threshold can explain many features of the experimentally observed dynamics \cite{gawthrop2011intermittent,milton2013intermittent}. However, much still remains unclear even in the case of relatively simple control tasks, such as real \cite{balasubramaniam2013control,cabrera2002onoff,milton2009balancing} or virtual \cite{loram2011human,foo2000functional,bormann2004visuomotor,loram2009visual} stick balancing. For instance, the mechanism behind extreme fluctuations of the systems under human control (resulting, e.g. in stick falls) still has to be explained \cite{cabrera2012stick}. 

Recently, a novel concept of noise-driven control activation has been proposed as a more advanced alternative to the conventional threshold-driven activation \cite{zgonnikov2014Inerface}. It argues that the activation of human control may be not threshold-driven, but instead intrinsically stochastic, noise-driven, and stem from stochastic interplay between operator's need to keep the controlled system near the goal state, on one hand, and the tendency to postpone interrupting the system dynamics, on the other hand. To justify the noise-driven activation concept  a novel experimental paradigm: balancing an overdamped inverted pendulum was employed~\cite{zgonnikov2014Inerface}. The overdamping eliminates the effects of inertia and, therefore, reduces the dimensionality of the system. Arguably, the fundamental properties and mechanisms of human control are more likely to clearly manifest themselves in such simplified setups rather than in more complicated conventional experimental paradigms. In the frameworks of human intermittent control with the noise-driven action the transition from passive to active phases is considered to be probabilistic, which reflects human perception uncertainty and fuzzy evaluation of the current state of the system  before correcting its dynamics.

The following two features revealed in these experiments are worthy of special attention. First, the phase space of system dynamics has to be extended in order to describe human actions during the active phase governed by the open-loop control mechanism. Namely, the cart velocity, a system parameter controlled \textit{directly} by a human operator, has to be treated as an individual phase variable. Second, the distribution function of the car velocity contains a sharp peak at the origin caused by the presence of the passive phase in human actions. This peak is the characteristic property of human intermittent control and seems to be a general feature which must be observed for various systems stabilized by human actions.

Driving a car in following a lead car is a characteristic example of human control. It allows us to hypothesize that the intermittency of human control should be pronounced in the driver behavior and affect the car motion dynamics essentially. The general objective of the present work is to verity this hypothesis. Its another purpose is to understand what factors are responsible for the found anomalous characteristics of the real traffic flow, at least, some of them. There are two principally different classes of such factors, external ones, in particular, various heterogeneities of road structure, and internal factors similar to the basic human properties noted above. The use of car driving simulators could enable us to separate their contributions due to the virtual environment being controllable completely (see, e.g., \cite{fisher2011handbook}).

By now a large amount of empirical and experimental data were collected about car following in the reality (see, e.g., \cite{de2012modelling}). In particular,  papers~\cite{wagner2003empirical,fouladvand2005statistical,wagner2006human,wagner2011time,wagner2012analyzing} have analyzed the data recorded by RTK-DGPS (real-time kinetics, differential global positioning system) on a Japanese test track \cite{gurusinghe2002multiple} and the single car data from the German freeway A1~\cite{knospe2002single}. In particular, the headway distribution $P(h)$, the relative velocity distribution $P(u)$, and the acceleration distribution $P(a)$ are used in this analysis. So in the given paper we also present these distributions constructed based on our data. 

\section{Car Driving Simulator and Experimental Setup}

\begin{figure}[h]
	\begin{center}
	  	\includegraphics[width=0.9\columnwidth]{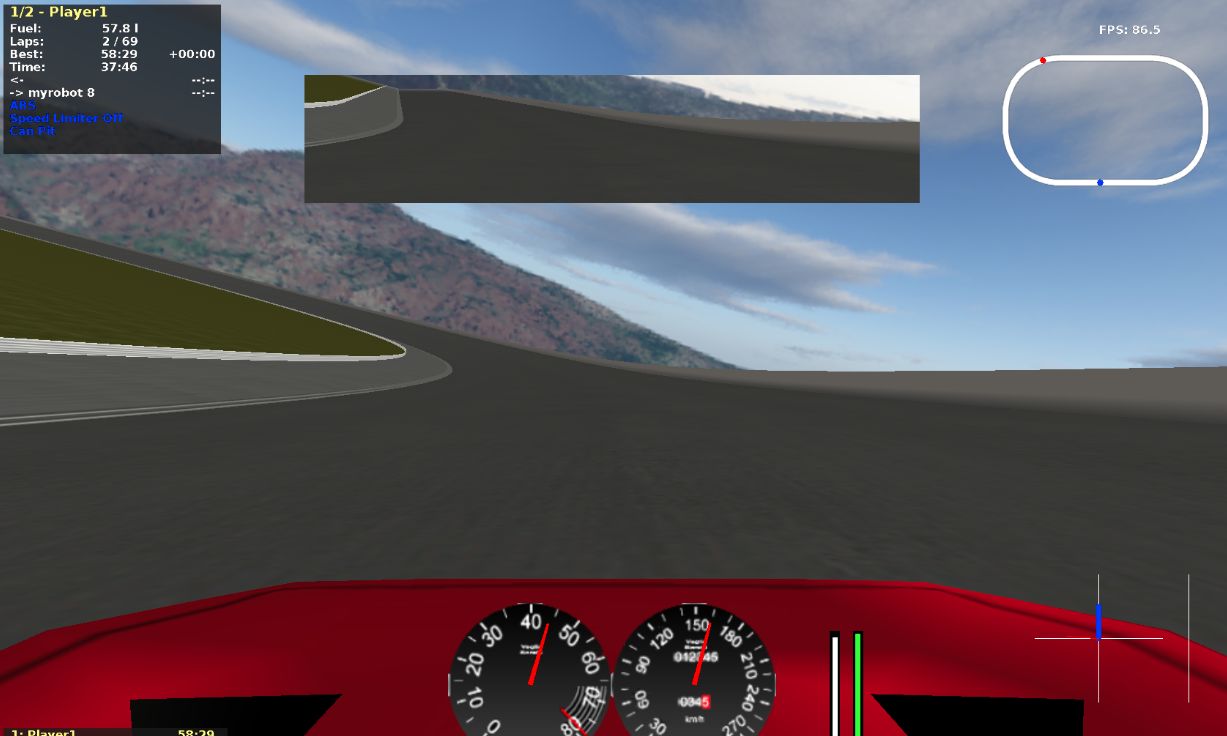}
	\end{center}
 	\caption{Car-following setup.}
 	\label{fig:one}
\end{figure}
		
A fair simple car-driving simulator was created using the open source engine TORCS (The Open Racing Car Simulator). It is a highly portable multi platform car racing engine widely used in ordinary and AI car racing games as well as a research platform \cite{TORCS_OffisSite2014}.	A commercially available high-precision steering wheel (Logitech G27 Racing Wheel) was used in the experiments. In the pilot experiments presented in this paper we used the available rectangular track (with smooth corners) of length about 3700~m and width of 30~m. A screenshot in Fig.~\ref{fig:one} illustrates the typical situation in the car-following experiments and the track geometry.  To keep up the human perception of the speed of motion through this environment the car moving ahead, the lead car, was controlled by computer such that its speed be kept rather close to 45.5~m/s (about 160 km/h, typical speed of cars in TORCS racing games). The individual data of the car positions and velocities were recorded at frequency about 50 Hz.  

Four male students of age about 22 participated in these experiments and each session was conducted twice on different days. The information on the subject's driving experience is given in Table~\ref{Tab1}, where the subjects are ordered according to their skill in driving real cars. The subjects were instructed (\textit{i}) to follow the lead car without overtaking it, (\textit{ii}) to keep a certain headway distance in order not to lose sight of the lead car, and (\textit{iii}) to continue the car driving for 30 minutes.

\begin{table}[h]
	\caption{The subject driving experience.}\label{Tab1}
	\begin{center}
		\begin{tabular} {|c|l|l|l|} \hline
		ID  & \textit{Driving License, issued} & \textit{Driving Periodicity} \\ \hline
		1   & Yes,  3 years ago   & Thrice a week \\ \hline
		2   & Yes,  4 years ago   & Once a week \\ \hline
		3   & Yes,  10 months ago & Rarely \\ \hline
		4   & No                  & None \\ \hline
		\end{tabular}
	\end{center}
\end{table}

\section{Results and Discussion}

For the reasons discussed in Introduction we had the hypothesis that, first, a multi-dimensional extended phase space would be required to describe the dynamics of a car driven by a human driver. Second, the distribution of the corresponding control parameter governed directly by human actions should contain a sharp peak at the origin. A driver controls the car dynamics via changing the position of the gas or break pedal, so the car acceleration $a$ and its time variation rate, called the jerk $j=da/dt$, may be additional phase variables. Therefore the recorded data were used to construct phase portraits and distribution functions for the four variables: the position of the car driven by a subject or, speaking more strictly, its headway distance $h$ (the distance between the lead and following cars), the car velocity $v$ and acceleration $a$, as well as the jerk $j = da/dt$. The car position and velocity were recorded directly by the simulator, whereas the acceleration and jerk were calculated using the Savitzky-Golay filters. 

\newcommand{\plotwidth}{\columnwidth} 
\newcommand{\plothight}{0.65\plotwidth}
		\begin{figure*}[p]
			\begin{center}
				 \centering{\scriptsize Subject 1 \hphantom{xxxxxxxxxxxxxxxxxxxxxxxxxxxxxxxxvvvvvvvvvccc}  Subject 4}
\\	
				\includegraphics[width = \plotwidth,height = \plothight]{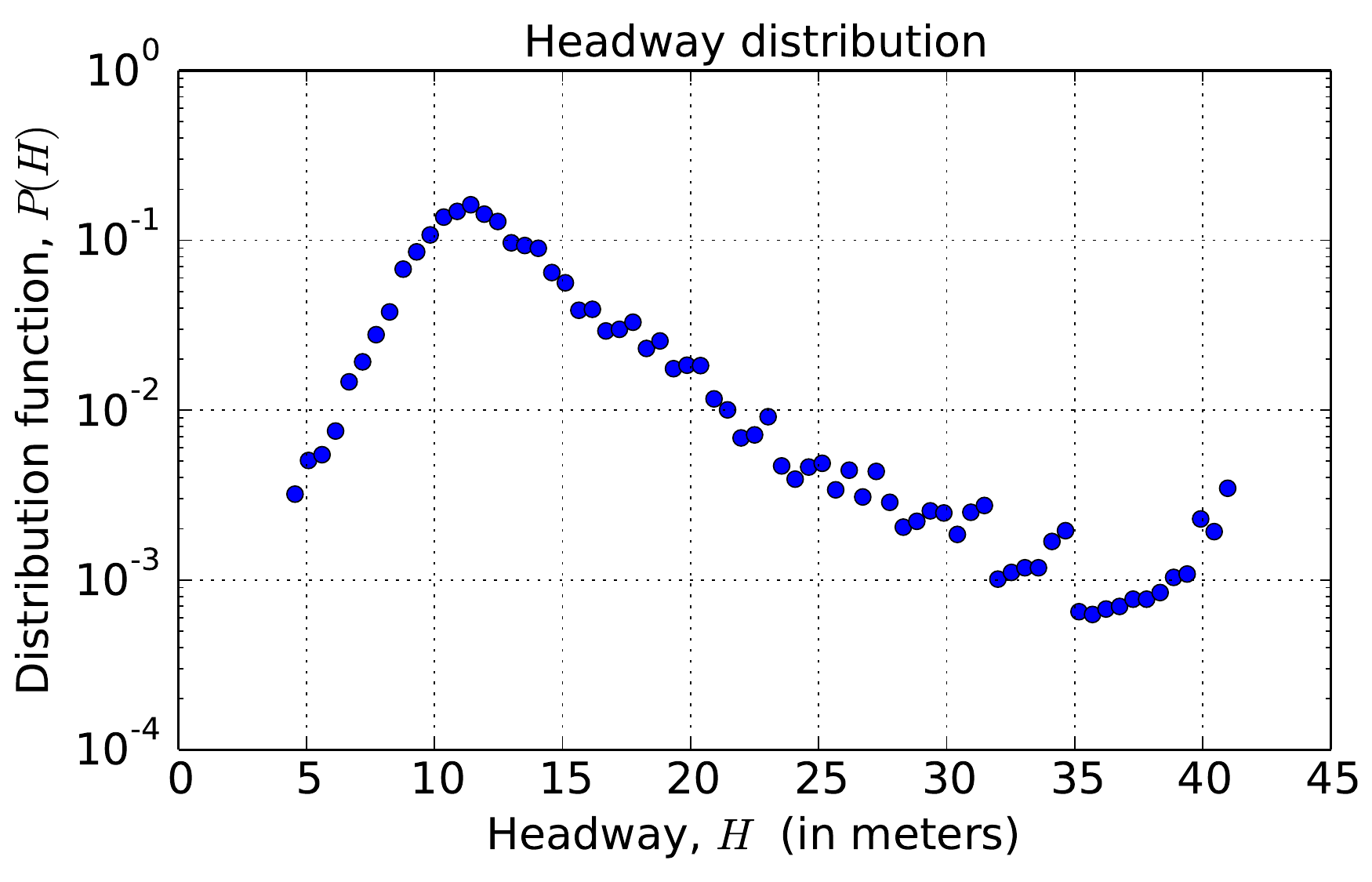}\quad
				\includegraphics[width = \plotwidth,height = \plothight]{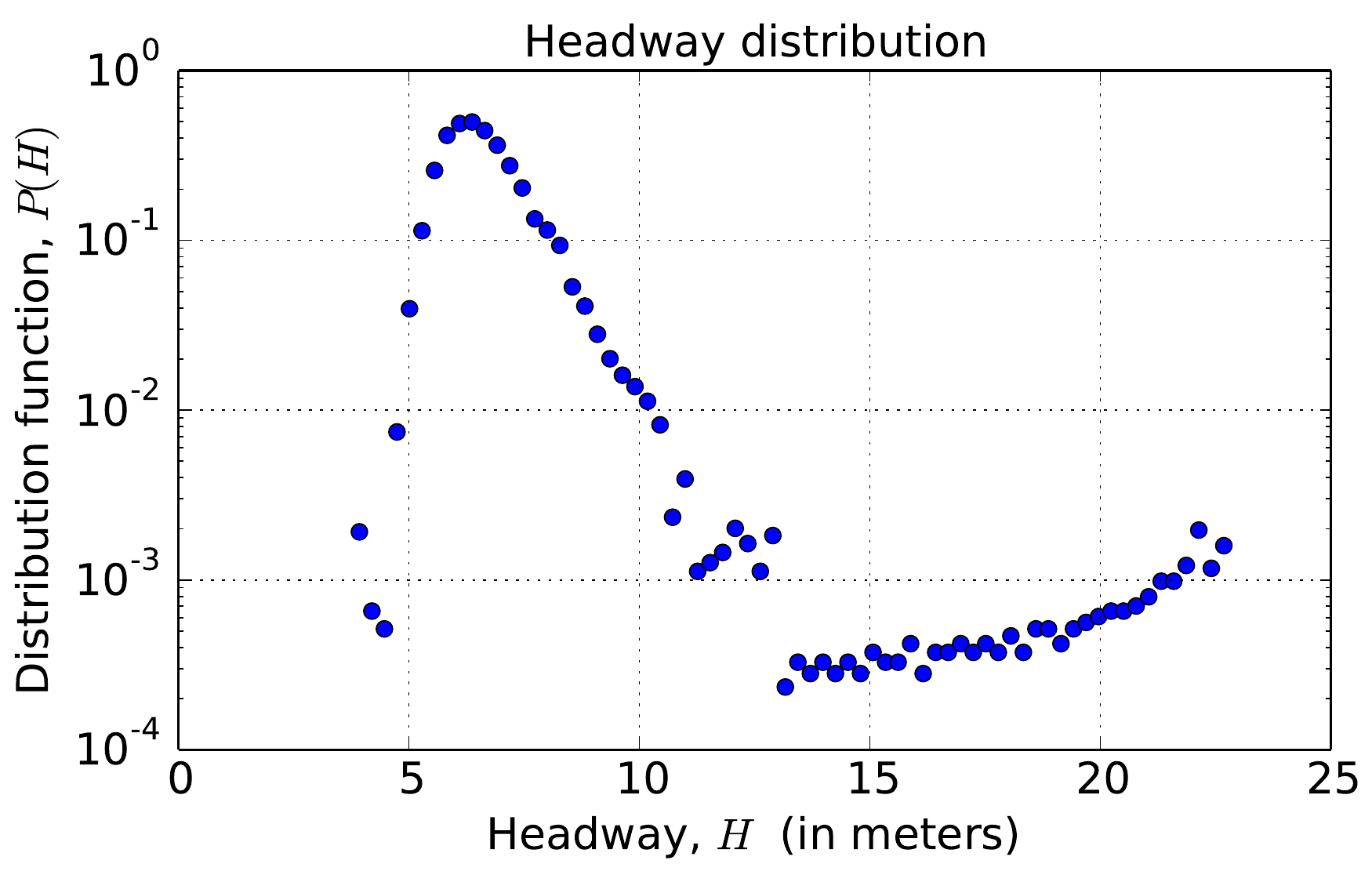}
\\
				\includegraphics[height = \plothight]{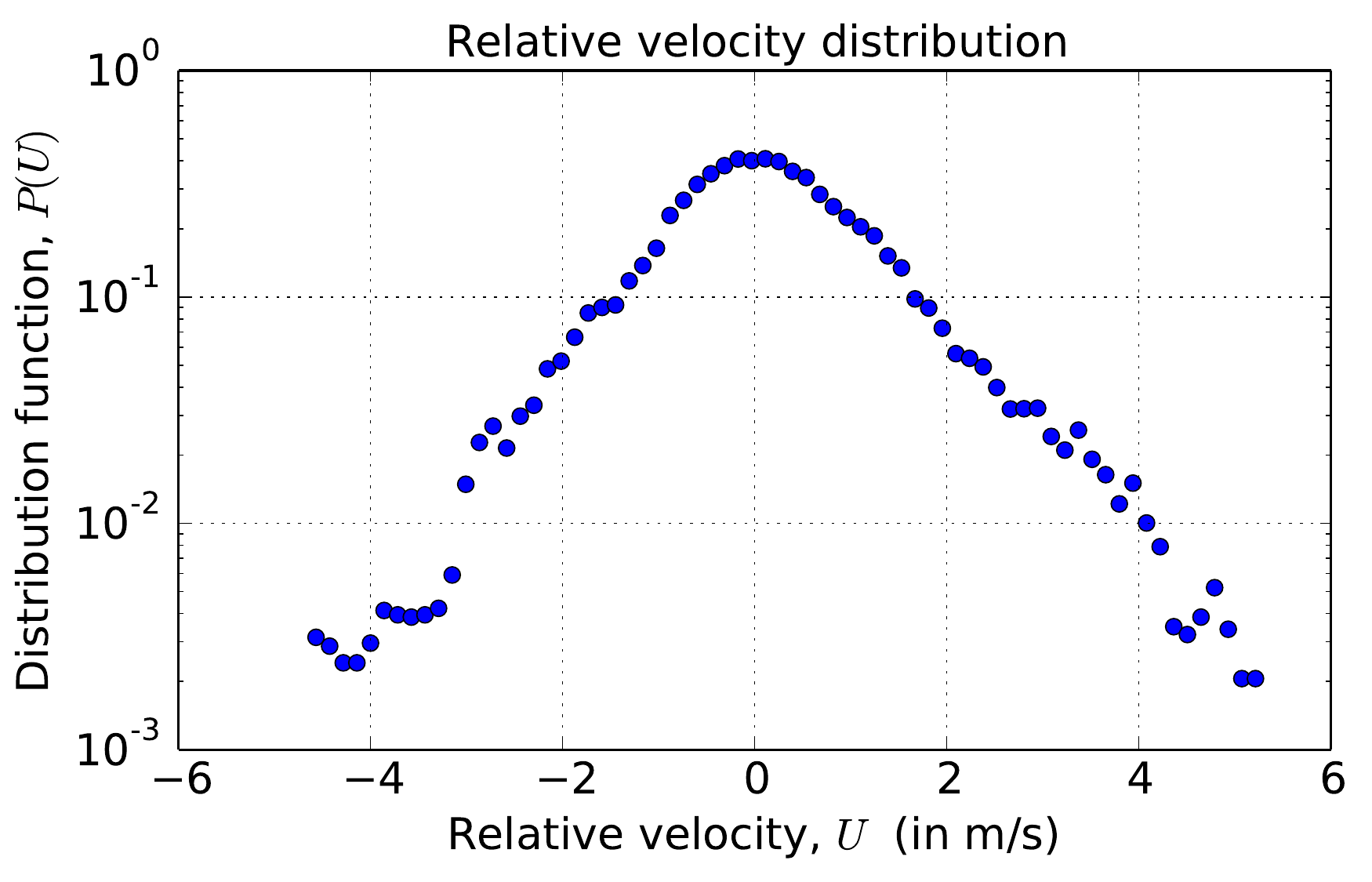}\quad
				\includegraphics[height = \plothight]{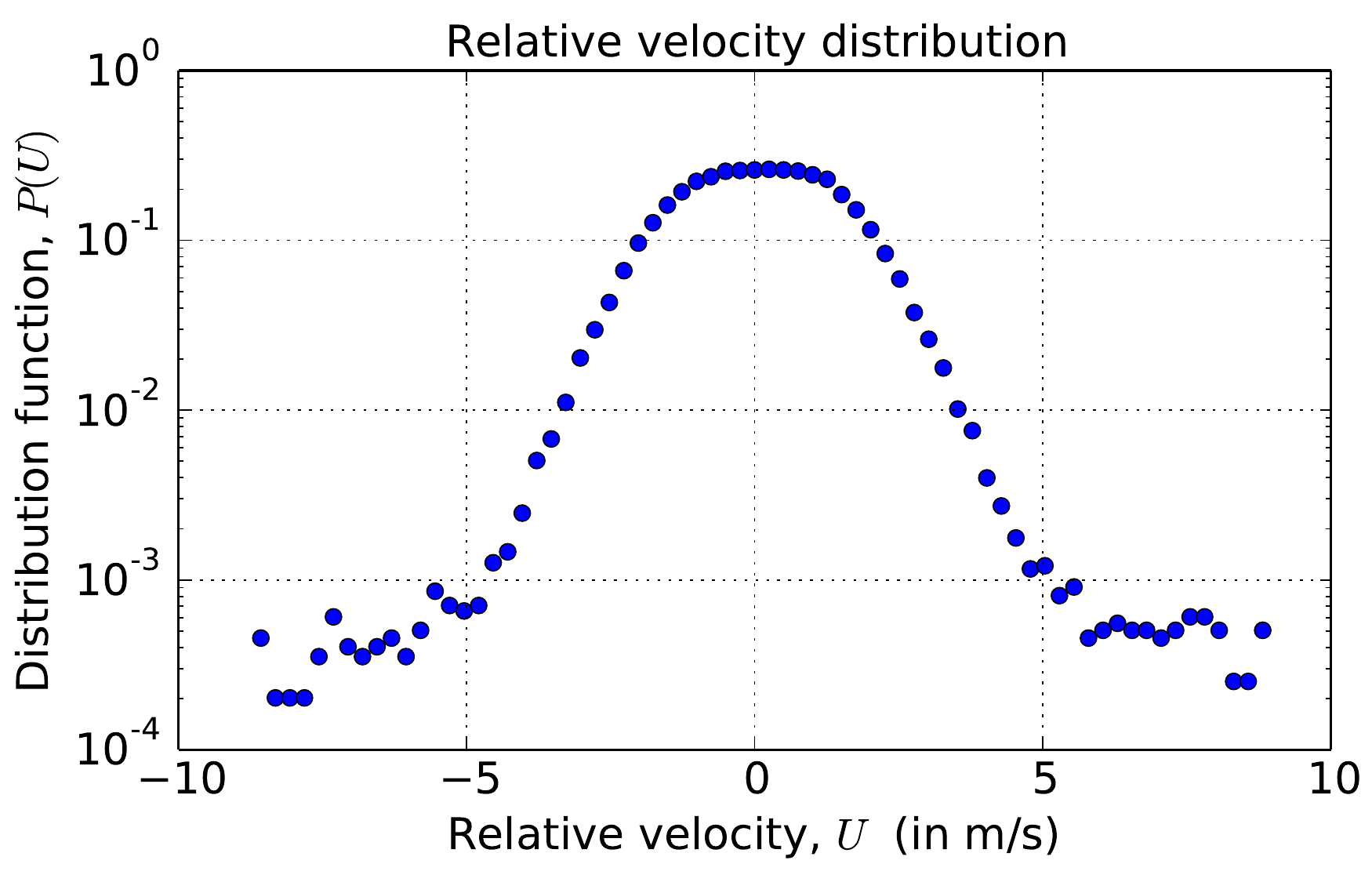}
\\ 
				\includegraphics[height = \plothight]{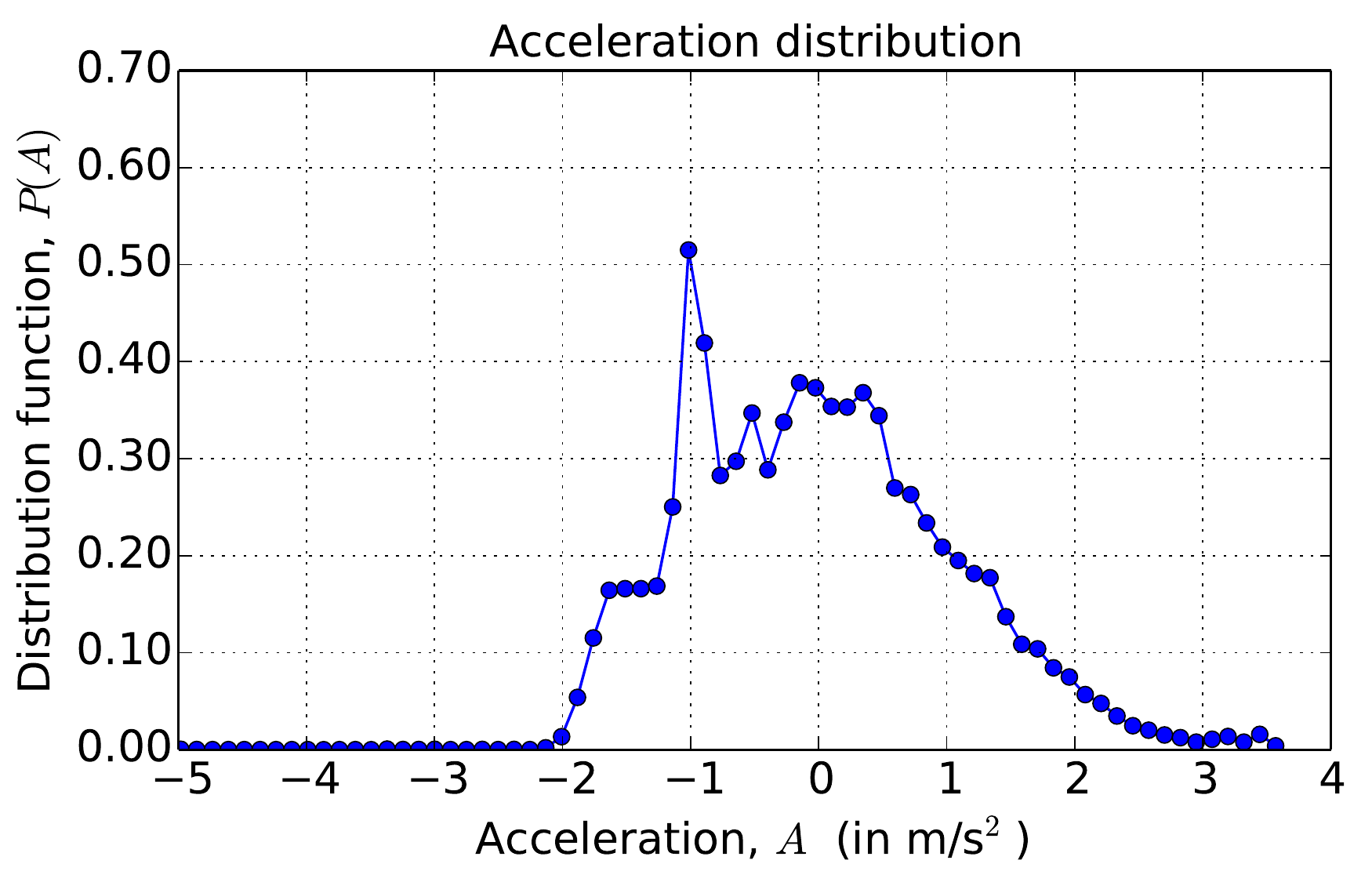}\quad
				\includegraphics[height = \plothight]{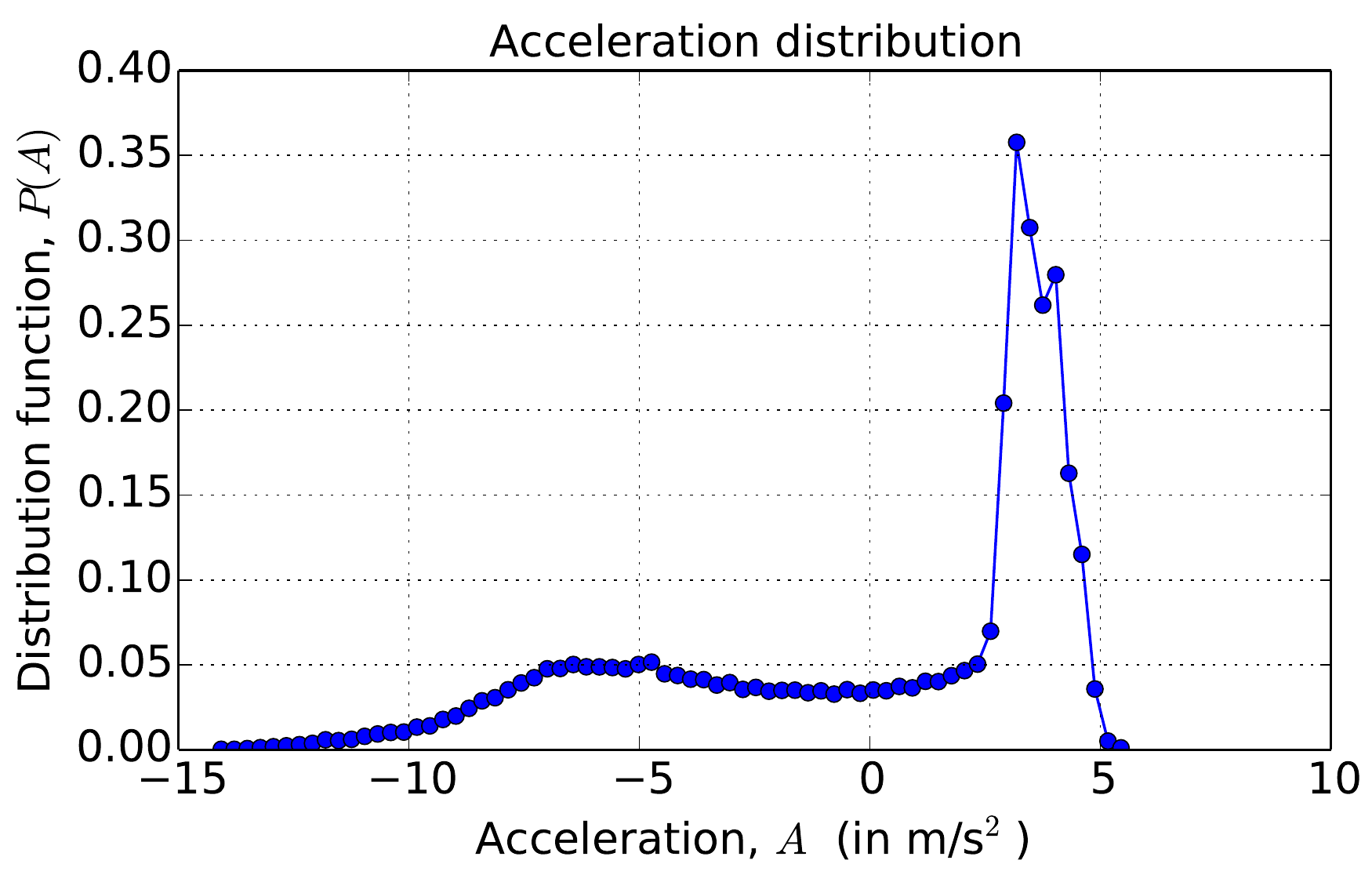}
\\
				\includegraphics[height = \plothight]{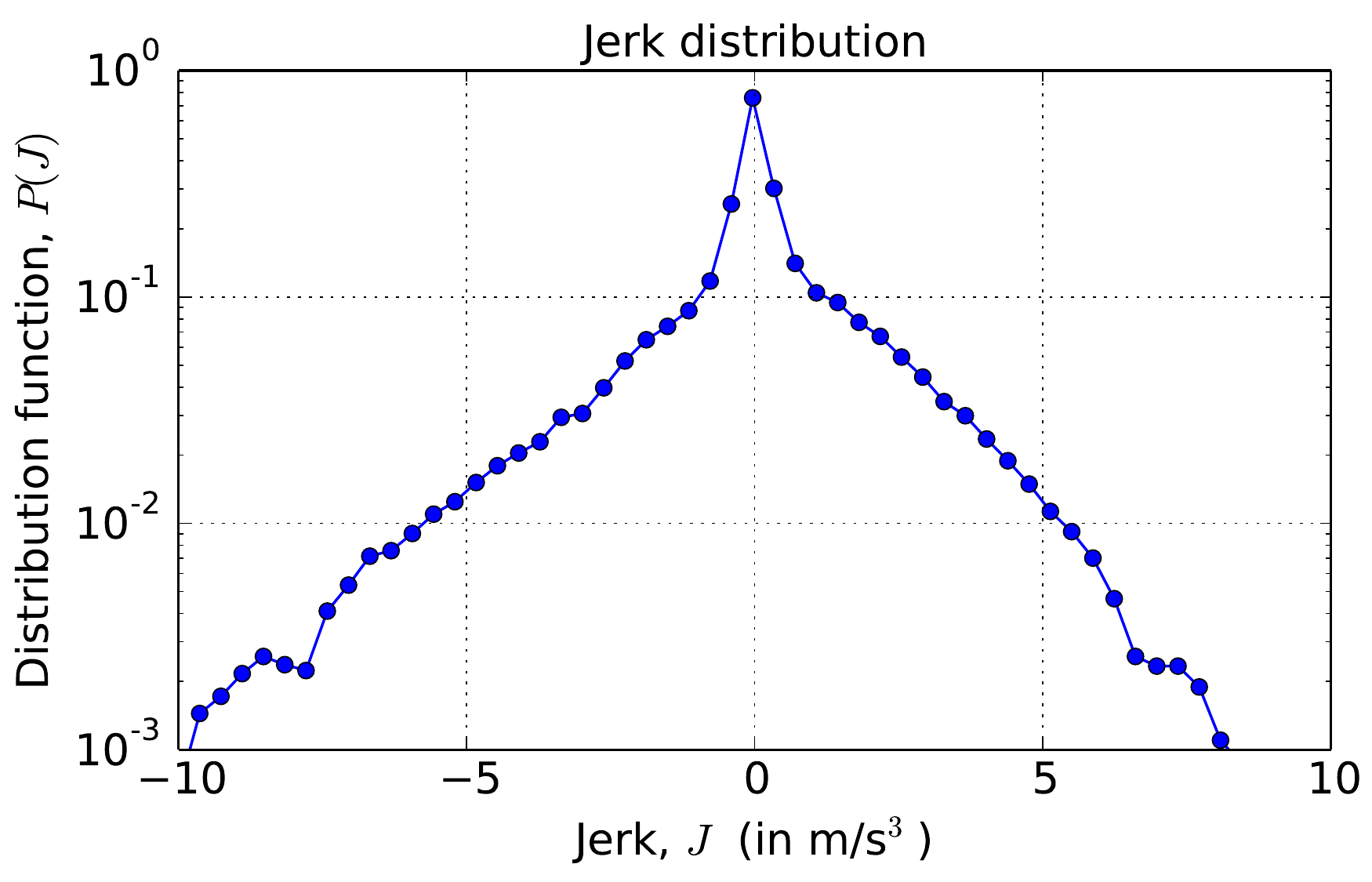}\quad
				\includegraphics[height = \plothight]{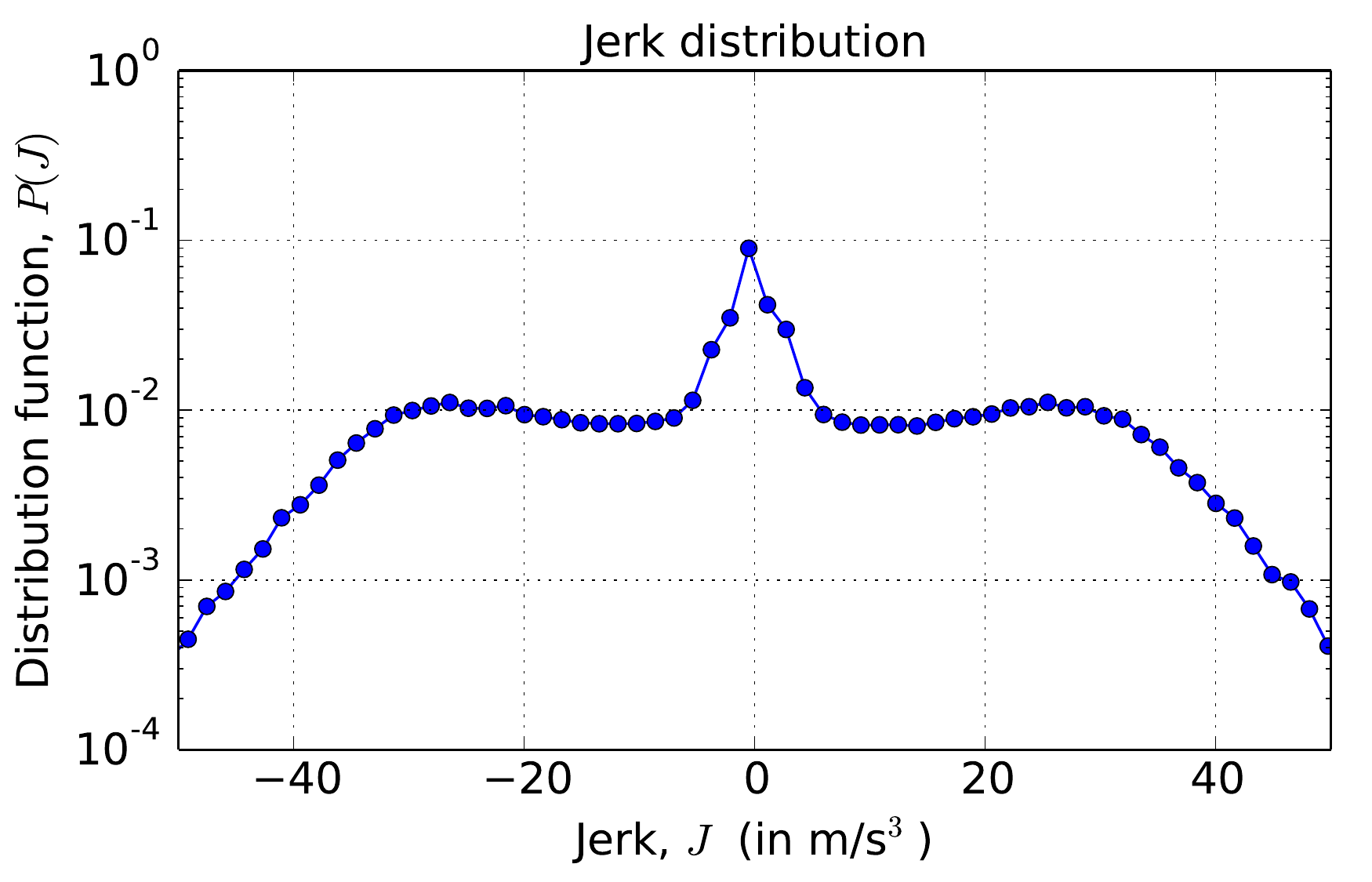}
				\caption{The distributions of headway distances, relative velocity, acceleration, and jerk constructed using the data collected by Subject~1 experienced in driving real cars (left column) and Subject~4 with no driving experience. }
				\label{fig:Res1}
			\end{center}
		\end{figure*}

In this work we confine the description of the obtained results to presenting the found distribution functions illustrated in Fig.~\ref{fig:Res1}. Two subjects, Subject 1 (left column) and Subject 4 (right column) are chosen to emphasize a possible dependence of the distribution functions on the driving skill. Figure~\ref{fig:Res1} depicts the constructed distribution functions of the headway distance $h$, the relative car velocity $u=V -v$, the acceleration $a$ and jerk $j$. The found distributions for the remaining two subjects with some experience in driving real cars are rather similar to that of Subject 1. These distribution functions are just histograms of the corresponding experimental records with the bin size chosen in such way that these histograms be maximally smooth without loss of information. 

Especially for Subject 1 the obtained distribution functions of the headway distance $P(h)$  and the relative velocity $P(u)$  turn out to be rather similar in shape and, partly, scales to the corresponding distributions functions found for the real traffic flow and in experiments with real car motion (cf., e.g., \cite{wagner2003empirical}). 

In particular, the shown forms of $P(h)$ demonstrate that the headway distance is distributed according to the asymmetric Laplace law; the experience of subjects is reflected mainly in the scales. In the conducted experiments the higher the experience of a subject, the larger headway distances that are kept in driving.  In literature (see, e.g., \cite{kesting2012traffic}) the gamma distribution is widely used to fit the corresponding experimental or empirical data. However, the found form of $P(h)$ is described by the gamma distribution only for large value of $h$. In the region of small values their behavior is different, which is worthy of special consideration. It should be pointed out that in the conducted experiments the quantity $\tau=h/V$, called the ``time-to-collision,''  takes extremely low values, for example, at the maximum of the headway distribution for Subject~1 it is about $\tau_m\sim 0.2$~s. The corresponding values in the real traffic are $\tau_m\gtrsim 1$~s. We explain these anomalously low values of the time-to-collision noting that, first, the human perception of distances in the reality and the virtual TORCS environment may be different. Second, in the given experimental setup the lead car moved at the fixed velocity, so for the subjects it was necessary to keep only the relative velocity rather low. It is justified by the fact that the relative velocity distributions obtained in these experiments and for the real traffic turn out to be practically of the same thickness. The effects of unexpected variations in the lead car velocity should be a subject of individual investigation.

The distribution functions of the acceleration and jerk (two lower rows in Fig.~\ref{fig:Res1}) depend substantially on the subject's individuality and reflect their personal styles of driving which seem to be rather similar for the three subjects with some driving experience. Nevertheless, as a general feature, we note the bimodal form of the acceleration distribution and a certain rather sharp peak all the jerk distributions possess at the origin, $j=0$. For the real traffic data the acceleration distribution is also of bimodal form \cite{wagner2003empirical}.

Appealing to the results of the experiments on balancing overdamped virtual pendulum \cite{zgonnikov2014Inerface} we can infer the following from the existence of these peaks. First, such a peak signals that there can be singled out considerable fragments of car motion when the acceleration was fixed. During the other fragments a driver seems to correct actively the car motion trying to change the current regime. This style of car driving can be categorized as the intermittent control being a sequence of alternate phases of active and passive behavior. So the obtained results may be regarded as a certain evidence for driver intermittent behavior.  Second, as a general property of human intermittent control, the parameter controlled \textit{directly} by a human operator in governing a mechanical system must be distributed according to a law with a central sharp peak. Besides, this control parameter has to be treated as an independent phase variable. Keeping in mind the obtained results we pose a hypothesis that the car dynamics, at least, in car-following has to be described using four phase variables, namely, the car position on the road (or headway distance), the car velocity, acceleration, and jerk. It should be noted that the first two variables meet the standard paradigm of Newtonian mechanics, whereas the last two variables are due to the human factor.

\section{Four-Variable Model of Car-Following}

\begin{figure*}[t]
\begin{center}
\includegraphics[width=0.9\textwidth]{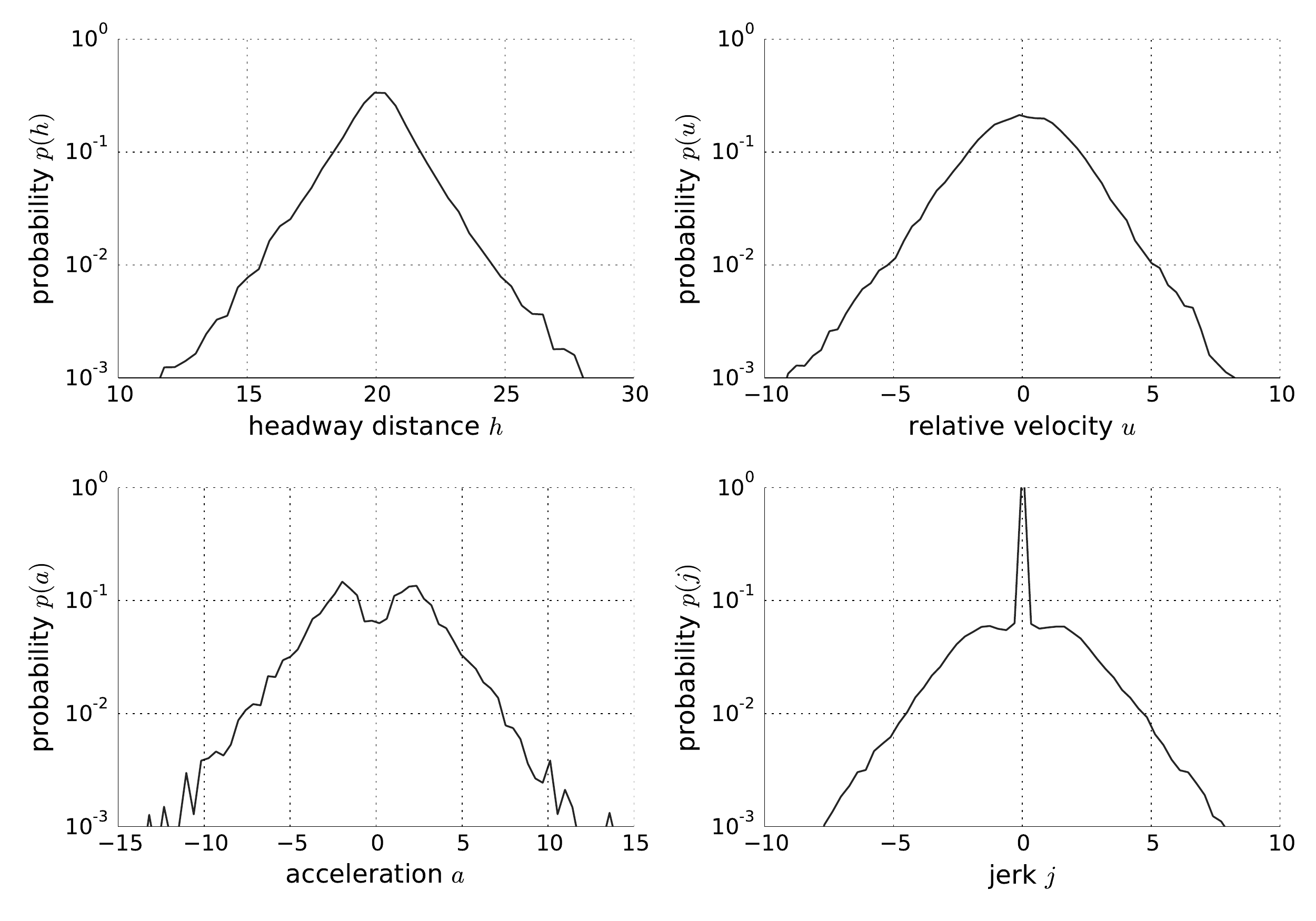}
\end{center}
\caption{The distributions of the headway distance $h$, relative velocity $u=v-V$, acceleration $a$, and the jerk $j\propto (\theta-v)$ obtained by numerical solution of  model~\eqref{m:1}--\eqref{m:4}. In simulation the following values were used, $v_\text{max} = 30$~m/s (about 100~km/h), $V = 15$~m/s, $D=20$~m, $a_\text{th}=0.1$~m/s$^2$, $\tau_h$ = $\tau_\theta = 0.2$~s, $\tau_v=1$~s, and $\epsilon=0.005$~m/s$^{1.5}$. The numerical labels at axes are given in the corresponding units composed of meters and seconds.}
\label{F:M}
\end{figure*}

In this section we discuss a mathematical model for car-following that employs the results of the experiments reported above. It is based on the assumption that to describe the driver behavior the extended phase comprising four independent variables is required. A driver is not able to change the car position and its velocity directly; he can only vary the car acceleration by pressing the gas or break pedal. In real driving the car acceleration on its own is an important characteristics of car motion. Therefore, in describing the car dynamics we have to include the car acceleration in the list of the phase variables \cite{lubashevsky2003rational,lubashevsky2003bounded,zgonnikov2014extended}. However, according to the found characteristics of the jerk distribution, \textit{the jerk on its own is also an independent phase variable} or another additional variable combining the headway distance $h$, the car velocity $v$, acceleration $a$, and jerk $j=da/dt$ within a certain relationship should be introduced. In the proposed model using a simplified description of car motion control, this fourth variable is the position $\theta$ of an effective pedal combining the gas and break pedals into one control unit.

Namely, the model is specified as follows. The car ahead is assumed to move at a fixed velocity $V$ and the dynamics of the following car is given by the equations

\begin{eqnarray}
\label{m:1}
\frac{dh}{dt} & = & V-v\,,\\
\label{m:2}
\frac{dv}{dt} & = & a\,, \\
\label{m:3}
\tau_{\theta}\frac{da}{dt} & = & \theta-a\,, \\
\label{m:4}
\tau_{h}\frac{d\theta}{dt} & = & \Omega\left(a-\theta \right)\cdot\left[a_{\text{opt}}(h,v)-a\right] +\epsilon\xi(t)\,.
\end{eqnarray}
Equations~\eqref{m:1} and \eqref{m:2} are just simple kinematic relations between the variables $h$, $v$, and $a$, equation~\eqref{m:3} describes the mechanical properties of the car engine and its response with some delay $\tau_\theta$ to the position $\theta$ of the control unit measured here in units of acceleration. The last equation~\eqref{m:4} describes the driver behavior. It combines the basic ideas of noise-driven activation in human intermittent control \cite{zgonnikov2014Inerface} and the concept of action dynamical traps for systems with inertia \cite{zgonnikov2014extended}. The driver is able to control directly only the position $\theta$ of the control unit and the difference $(\theta - a)$ between the desired acceleration $\theta$ and the current car acceleration $a$ is the parameter quantifying the difference between his active and passive behavior. The bounded capacity of driver cognition is described in terms of action dynamical traps via the introduction of cofactor         
\begin{equation}
\Omega(a-\theta)  = \frac{(a-\theta)^{2}}{(a-\theta)^{2}+a_{\text{th}}^{2}}
\end{equation}
similar to fuzzy reaction coefficients. Here $a_\textit{th}$ is the driver perception threshold of car acceleration.  The ansatz
\begin{equation}
a_{\text{opt}}(h,v)=\frac{1}{\tau_{v}}\left[v_{\text{max}}\frac{h^{2}}{h^{2}+D^{2}}-v\right]
\end{equation}
determines the optimal acceleration with which the strictly rational driver with perfect perception would drive the car. This expression inherits the optimal velocity mode widely used in modeling traffic flow (see, e.g.. \cite{kesting2012traffic}). Here $\tau_v$ is the human response delay time, $v_\text{max}$ is the maximal velocity acceptable for safety reasons on a given road without neighboring cars, and $D$ is the characteristic headway distance when drivers consider it necessary to slow their cars down as the headway distance decreases. The last term in equation~\eqref{m:4} is the random Langevin force, where $\xi(t)$ is white noise of unit amplitude and $\epsilon$ is the Langevin force intensity. The interplay between the fuzzy perception function $\Omega(\theta-a)$ and this Langevin force are two main components of the noise-induced \{\}activation model elaborated in \cite{zgonnikov2014Inerface} for describing human balance of overdamped pendulum.  Finally, the difference $[a_{\text{opt}}(h,v)-a]$ quantifies the stimulus for the driver to correct the current state of car motion.

In the given paper we present a preliminary investigation of this model and the goal of this section is to demonstrate a potential capability of such an approach to describing complex properties of real traffic flow. Figure~\ref{F:M} depicts the results of numerical solution of model~\eqref{m:1}--\eqref{m:4} using the characteristic values of the systems parameters employed by other models, at least, being of the same order (cf., e.g., \cite{kesting2012traffic}). It should be noted that the distributions obtained by numerical simulation of the developed model and constructed based on the experimental data collected by Subject~1 look rather similar.  

\section{Conclusion} 

The conducted experiments employing the TORCS car driving simulator have demonstrated that the behavior of subjects involved into driving virtual cars should be categorized as the generalized intermittent control over mechanical systems. It consists of a sequence of alternate fragments of active and passive phases of driver behavior. The passive phase is characterized by the fact that during the corresponding time interval a driver does not change the position of the gas or break pedal. In this case the jerk plays the role of the parameter controlled directly by the driver and, so, has to be regarded as an independent phase variable determining the car dynamics. It enabled us, keeping on mind also driving real cars, to hypothesize that a sophisticated description of car motion controlled by human actions requires the introduction of four dimensional phase space, where the car position, velocity, acceleration, jerk are the independent variables. 

A new model for the car-following that allows for these features has been proposed. Its numerical simulation has demonstrated that the combination of the concepts of the noise-driven activation in human intermittent control and the action dynamical traps caused by the bounded capacity of human cognition can reproduce, at least, qualitative the results of the conducted experiments.

\section*{Acknowledgments}

This work was supported in part by the JSPS Grants-in-Aid for Scientific Research Program, Grant 24540410-0001.


\end{document}